\documentstyle[emulateapj,apjfonts,epsf]{article}

\def\Halpha{H$\alpha$~}
\def\about{$\sim$}
\def\arcsec{$\,^{\prime\prime}$~}
\def\arcmin{$\,^\prime$~}
\def\deg{$^{\circ}$~}
\def\erg/cm2sec{ergs~cm$^{-2}$~s$^{-1}$}  
\def\ergcm2{ergs~cm$^{-2}$}

\def\X{$\times$~}

\def\Lx{L$_x$~}

\def\pc3{pc$^{-3}$~}

\def\cm-3{cm{$^{-3}$}~} 
\def\km/s{km~s$^{-1}$~}

\newcommand{\lsim }{{\lower0.8ex\hbox{$\buildrel <\over\sim$}}}
\newcommand{\gsim }{{\lower0.8ex\hbox{$\buildrel >\over\sim$}}}

\newcommand{\Msun}{\ifmmode {M_{\odot}}\else${M_{\odot}}$\fi~}
\newcommand{\Rsun}{\ifmmode {R_{\odot}}\else${R_{\odot}}$\fi}
\newcommand{\Lsun}{\ifmmode {L_{\odot}}\else${L_{\odot}}$\fi}
\newcommand{\mv}{\ifmmode {m_{V}}\else${m_{V}}$\fi}
\newcommand{\Mv}{\ifmmode {M_{V}}\else${M_{V}}$\fi}
\newcommand{\lopt}{\ifmmode L_{opt} \else $~L_{opt}$\fi}
\newcommand{\loglopt}{\ifmmode{\rm log}~L_{opt} \else log$~L_{opt}$\fi}
\newcommand{\lx}{\ifmmode L_x \else $~L_x$\fi}
\newcommand{\loglx}{\ifmmode{\rm log}~L_x \else log$~L_x$\fi}
\newcommand{\cmsq}{\ifmmode{\rm ~cm^{-2}} \else cm$^{-2}$\fi}
\newcommand{\nh}{\ifmmode{\rm N_{H}} \else N$_{H}$\fi}
\newcommand{\fcgs}{\ifmmode {\rm erg~cm}^{-2}~{\rm s}^{-1}\else
erg~cm$^{-2}$~s$^{-1}$\fi} 
\newcommand{\lcgs}{\ifmmode erg~~s^{-1}\else erg~s$^{-1}$\fi}

\received{}
\accepted{}
\begin{document}

\title{Chandra Exposes the Core Collapsed Globular Cluster NGC 6397}

\author{J. E. Grindlay, C.O. Heinke, P.D. Edmonds,
S.S. Murray\altaffilmark{1} and A.M. Cool\altaffilmark{2}}

\altaffiltext{1}{Harvard-Smithsonian Center for Astrophysics, 60 Garden St,
Cambridge, MA 02138; josh@cfa.harvard.edu; cheinke@cfa.harvard.edu;
pedmonds@cfa.harvard.edu; smurray@cfa.harvard.edu}

\altaffiltext{2}{Department of Physics and Astronomy, San Francisco State
University, 1600 Holloway Avenue, San Francisco, CA 94132; cool@sfsu.edu}


\begin{abstract}
We report results of the Chandra deep imaging observation 
of the closest post-core collapse globular cluster, NGC 6397. 
Some 25 sources are detected within 2\arcmin of the cluster 
center of which \about20 are likely cluster members  
with \Lx \gsim3 \X 10$^{29}$ \lcgs. The x-ray spectra 
suggest identifications with  1 quiescent low mass x-ray binary 
(qLMXB) detected by the thermal emission from its neutron 
star (NS) and 9 cataclysmic variables (CVs), 8 of which are 
identified in our deep HST imaging survey (reported separately). 
Three (of 16) BY Draconis main sequence binary candidates identified 
in our earlier HST imaging study (Taylor et al) 
are  detected of which one is indeed the counterpart of the eclipsing 
millisecond pulsar (MSP) recently located by D'Amico et al. 
Two other BY Dra candidates are also detected, whereas  
none of the probable He white dwarf (WD) binaries identified 
by Taylor et al are  indicating they do 
not contain MSP primaries. 
The x-ray color magnitude diagram suggests  
that the remaining 5 probable cluster sources are a mixture of 
faint CVs, BY Dra binaries and MSPs.  
Compact binaries containing WDs appear to dominate this cluster, 
in contrast to those containing NSs in 47Tuc.

\end{abstract}

\section{INTRODUCTION}
The remarkable high resolution x-ray imaging of the Chandra X-ray 
Observatory has opened a new era for the study of compact binaries 
in the dense cores of globular clusters. These systems 
have now been imaged in 47Tuc (Grindlay et al 2001; hereafter 
GHEM) and found to include neutron stars (both quiescent 
low mass x-ray binaries (qLMXBs) and  millisecond pulsars 
(MSPs)), white dwarfs (cataclysmic variables (CVs)), and main 
sequence binaries (BY Draconis stars). X-ray imaging, with    
spectroscopy and timing, enables particularly direct discovery 
and study of these systems which not only constrain stellar 
and binary evolution, but the very dynamical evolution of cluster 
cores. 

We report initial results for our Chandra observation 
of the closest (2.5kpc) core 
collapse globular, NGC 6397, in which we have 
previously discovered a population of faint x-ray sources (Cool et 
al 1993), some of which have been identified with 
CVs (Cool et al 1995, Grindlay et al 1995). 
Deeper ROSAT observations of the cluster have been reported by 
Verbunt and Johnston (2000). 
The Chandra observations, together with deep HST imaging studies 
(Taylor et al 2001, hereafter TGEC; Grindlay et al, 
in preparation; hereafter 
GTEC), are a factor of 30 more sensitive and much higher resolution 
and thus allow a nearly complete census of the compact binaries in 
this dynamically interesting globular. 

\section{CHANDRA OBSERVATIONS AND ANALYSIS}
As part of the GTO program, we obtained a 49ksec observation 
of NGC 6397 with ACIS-I starting at UT15:31 July 31, 2000,    
in timed exposure mode in a single continuous observation. 
The cluster center 
(Sosin 1997) was placed at 1\arcmin in from the I3 chip boundaries so that 
the full core could be observed without chip gap complications. 
Standard CIAO2.1 processing tools 
were used to reduce the data, and Wavdetect revealed 68 
sources within the central 4\arcmin radius field. For this 
paper we restrict analysis to the central 2\arcmin radius, 
in which 25 sources are detected in the medium band (0.5-4.5keV) 
above a threshold (95\% confidence) of 3 counts. Source locations 
and a color composite Chandra image of the central (\lsim1\arcmin) 
sources are shown in Figures 1a and b, respectively. 

Event files were extracted using the Wavdetect source regions,  
which include $\sim$ 94\% of the counts associated with each source.  
The number of counts in each of three bands, 
0.5-4.5 keV (medcts), 0.5-1.5 keV (softcts), and 1.5-6 keV (hardcts) were 
extracted from these events files.  The results are plotted as an instrumental 
x-ray color magnitude diagram (cf. GHEM) in Figure 2, and source 
positions, fluxes, spectral fits and probable identifications 
(see below) are given in Table 1. 

\subsection{X-ray Spectra and Emission Models}

For sources with medcts \gsim50, spectra were extracted in bins 78 eV wide
which were grouped for \gsim9 counts/bin. Background spectra, derived from
a 43 pixel radius including no sources, were subtracted and sources were
fit to blackbody (BB), power law (PL), thermal bremsstrahlung (TB), and
two-component BB and PL models with XSPEC. TB models were acceptable fits
($\chi^2_{\nu} <$ 1.5, null hypothesis probability $>$ 5\%) for all
sources, and are recorded in Table 1 for all sources except U24 (ROSAT
source B; see below).  BB models alone are excluded for five (U17, U18,
U19, U22, U23) of the seven brightest sources ($\chi^2_{\nu} >$ 1.5, prob
$<$ 5\%).  PL models are also acceptable fits ($\chi^2_{\nu} <$ 1.5, prob
$>$ 5\%) to each of the sources (except U24), with best-fitting photon
indices in the range 1.5-1.7 except for U12 (2.4$^{+1.1}_{-0.6}$), and U18
(1.0$^{+0.3}_{-0.1}$).  BB + PL models could be rejected as having
physically unrealistic small BB radii (\about10m). All spectral fits
include the absorption column \nh, which is significantly enhanced above
the cluster value (\nh = 1 \X 10$^{21}$ \cmsq) for source U10 (the blue
source in Fig. 1b) as well as most of the other CVs but not for the BY
candidate U18 (see below).

Source U24 is similar in x-ray color (Fig. 2) and spectral shape to the
quiescent low-mass x-ray binaries (or soft x-ray transients in
quiescence) in $\omega$ Centauri (Rutledge et al. 2001) and in 47 Tuc
(GHEM, and Heinke et al. 2001; hereafter HGLE).  A BB spectral fit
is acceptable (kT=0.19$\pm0.02$ keV, $N_H$=$5\pm5\times10^{20}$
cm$^{-2}$), as are power law fits (with a steep unrealistic photon 
index of 6$\pm1$ and high $N_H$, $5^{+2}_{-1}\times10^{21}$ cm$^{-2}$)
and thermal bremsstrahlung fits (with kT=0.33$\pm0.05$ keV,
$N_H$=$2.2\pm1\times10^{21}$ cm$^{-2}$).  None of these are
physically realistic models, and we turn to hydrogen or helium
atmosphere neutron star models (Zavlin, Pavlov, \& Shibanov 1996;
Rajagopal \& Romani 1996) physically motivated by radiation of heat
accumulated in the core of an accreting neutron star during transient 
episodes and reradiated during quiescence (Brown, Bildsten, \&
Rutledge 1998).  This model has been fit
successfully to the qLMXBs Cen X-4,
Aql X-1, CXOU 132619.7-472910.8 in $\omega$ Cen; and X5 and X7 in 47
Tuc (Rutledge et al. 2001 and references therein; HGLE).  

\vspace*{0.4cm}
\hspace*{-0.3cm}
\epsfxsize=8.5truecm
\epsfbox{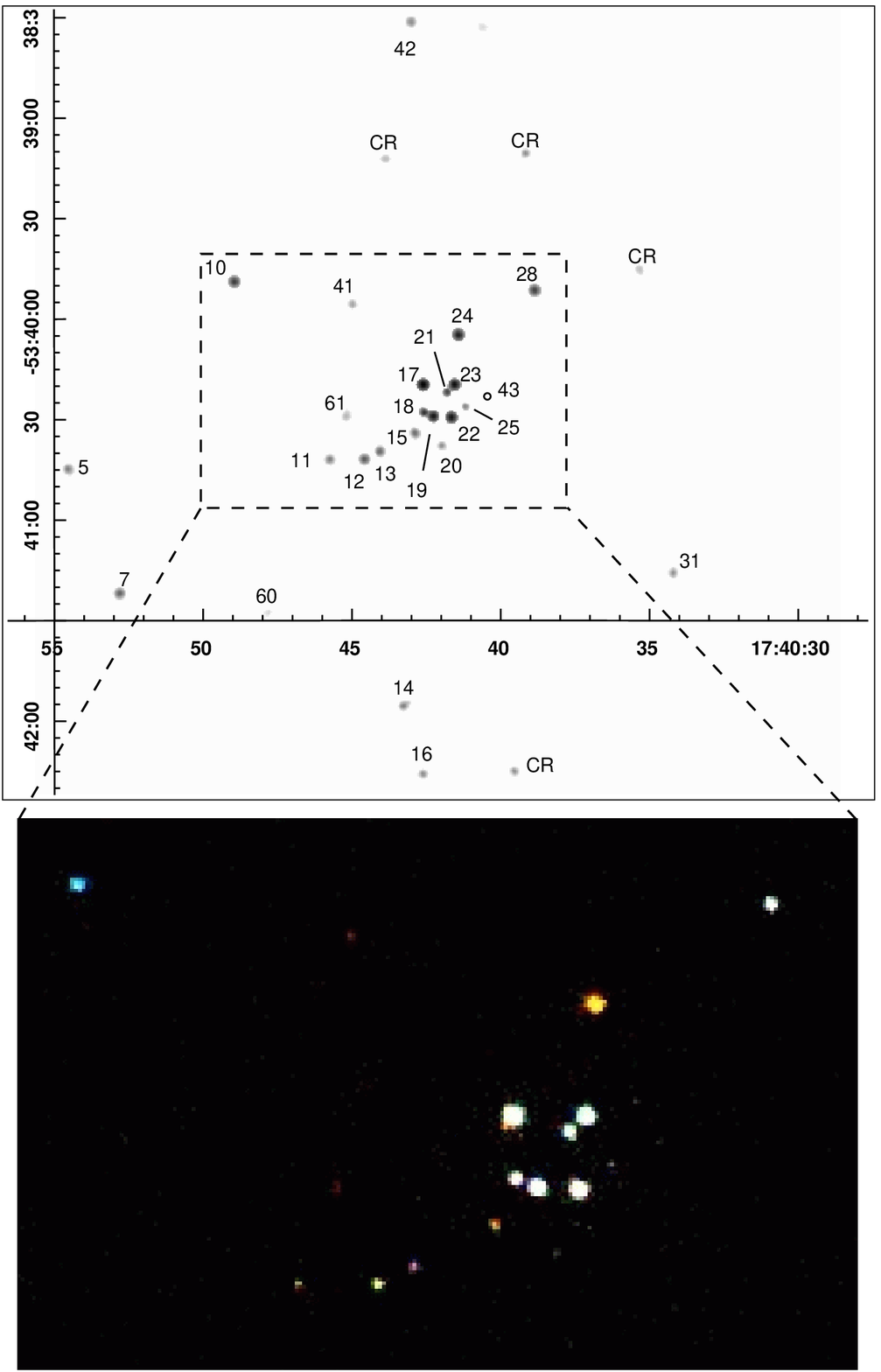}
\vspace*{-0.0cm}
\figcaption[f1.eps] 
{a) (top) Finding chart for sources detected with Wavdetect 
in medium band (0.5-4.5keV), with 4 probable cosmic ray (CR) events 
(all counts recorded within \lsim30sec) marked,  
and  b) (bottom) x-ray color image constructed 
from soft (0.5-1.2keV; red), medium (1.2-2keV; green) and hard 
(2-6keV; blue) bands for sources detected with Chandra in NGC 6397.
The cluster center is within \about1\arcsec\ of U19 (CV2).}
\bigskip

We fit the unmagnetized models of Lloyd \& Hernquist (2001, in
preparation), which assume a completely ionized atmosphere and opacity
due to coherent electron scattering and free-free absorption. 
A neutron star surface gravity of log $g_s$=14.38 was assumed, yielding a
gravitational redshift of 0.306.  Either a hydrogen or helium 
atmosphere gives a good fit (see Table 1 for kT$_{eff}$), with 
implied radii of the neutron star
(R$_{\infty}$, as seen at infinity) of 4.9$^{+14}_{-1}$ km for a hydrogen
atmosphere, and 12.0$^{+3}_{-7}$ km for a helium atmosphere.  These
are consistent with theoretical predictions for R$_{\infty}$ which span
the range 10 to 18 km (Lattimer \& Prakash, 2001), and the similarity
of the fits to better-constrained fits on X5 and X7 in 47Tuc suggests that 
we are indeed seeing thermal emission from the whole surface of the
neutron star. No power law component is seen, with only 1\% of the
total emission above 2.5 keV and only one photon above 3.3 keV. A
marginal (\about1$\sigma$) emission feature is seen in the spectrum at
0.9 keV, similar to the more significant features seen in X5 and X7 
in 47 Tuc.  The lack of variability and 
complete lack of a power law component suggest
that no accretion onto the neutron star surface is currently taking
place (see HGLE).

\vspace*{-0.1cm}
\hspace*{-0.7cm}
\epsfxsize=9.4truecm
\epsfbox{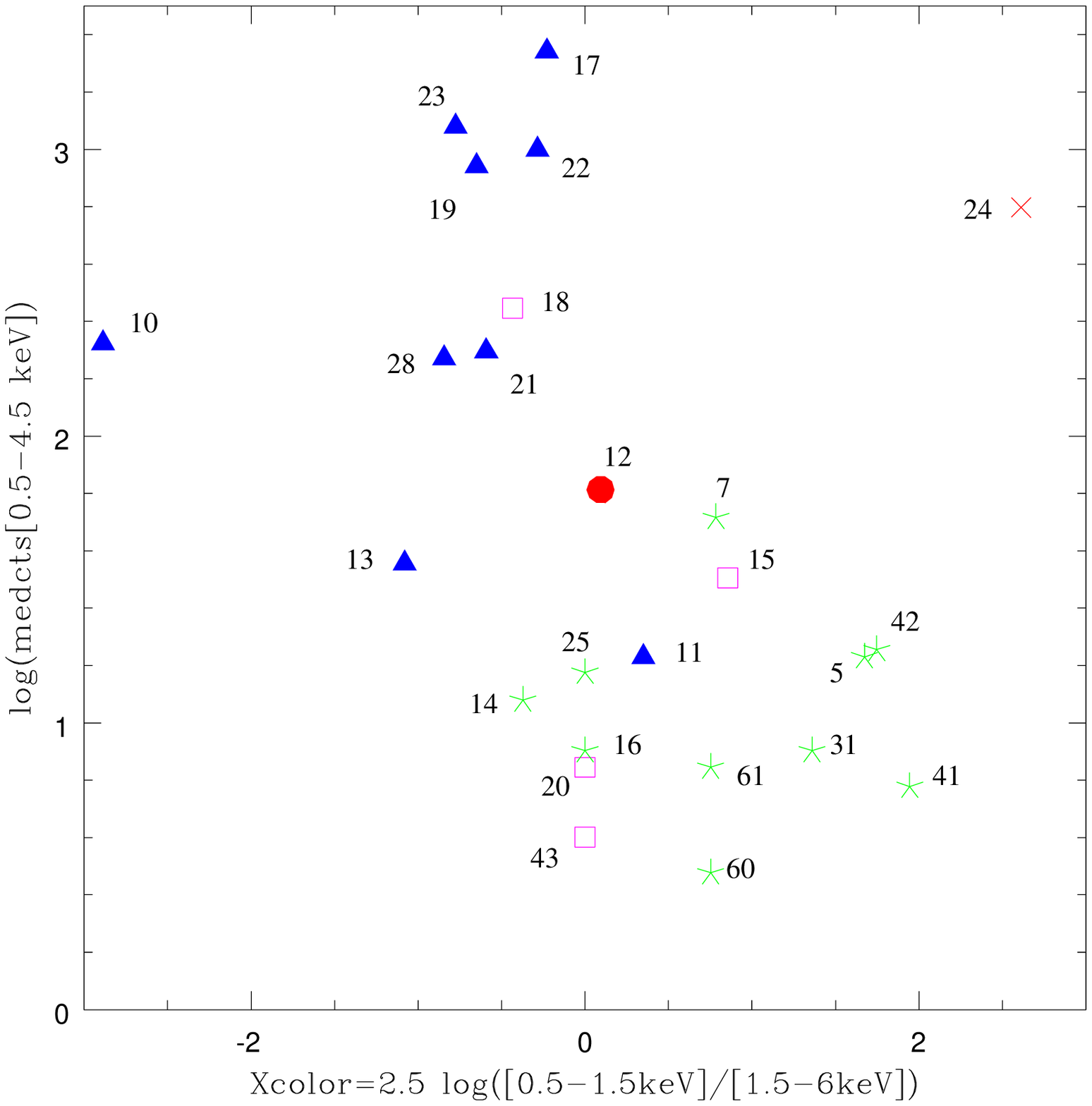}
\vspace*{-0.1cm}
\figcaption[f2.eps]
{X-ray color magnitude diagram, with probable source 
types labeled (as in GHEM): X = qLMXB, $\bullet$ = MSP,  
$\Delta$ = CV, $\Box$ = BY, 
$\ast$ = Unidentified. Source numbers correspond to those in Table 1.}
\bigskip

\subsection{X-ray Variability}
A dramatic eclipse (duration \about0.5hour) for source U23 (CV1)
is seen midway through the 49ksec 
observation, consistent with the 11.3 hour period discovered for this 
eclipsing CV (GTEC). Similarly, U17 (CV3) shows smooth
variations. Pulsation analysis for the CVs reveals several 
candidate periods, but longer observations (to be conducted 
in Chandra cycle 3) are required. 
The BY candidate U20 is clearly flaring whereas 
the surprisingly hard spectrum BY (or possible RS CVn) U18 appears 
constant as do U15 and U43. Source U12 shows a smooth sinusoidal 
like variation (discussed below) over the Chandra observation. 
 
\section{PROBABLE IDENTIFICATIONS}
The x-ray spectral results suggest 9 CVs, all with moderately 
hard TB spectra and internal self-absorption. The intrinsic \nh, 
particularly for U10 (CV6), suggests these systems may be dominated 
by magnetic CVs which show internal absorption from their 
``accretion curtains'' (cf. GHEM and references therein). 
Three of these CVs (CVs 1-3) were originally identified as such in 
our initial HST \Halpha survey (Cool et al 1995).  Two others (CVs 4-5)
were found via variability or as counterparts to ROSAT HRI sources 
(Cool et al 1998; Grindlay 1999).  CVs 6-8 were discovered in our 
deeper followup HST \Halpha survey (GTEC).  
The 9th ($=$ ROSAT source A) we identify as a CV on the basis of its 
Chandra spectrum, which closely resembles that of U21 (CV4); it lies
outside the HST field of view.

Using the very likely identifications with CVs 1-8, the required shifts in
RA, Dec between HST and Chandra are 1.24\arcsec, -0.03\arcsec, due
(primarily) to the difference in absolute astrometry of their respective
guide star systems (all positions in Table 1 are on the Chandra reference
frame, currently accurate to \about0.6\arcsec (1$\sigma$) (Aldcroft et al
2000). This solution allows precise (\lsim0.1\arcsec) positional searches
for the BY Dra stars identified (2 from TGEC, and 2 from subsequent
analysis). The HST detected CVs and BYs are marked in the CMDs of Figure 3.
No star is seen at the position of the likely qLXMB and we set a limit for
the optical companion of $M_V > 11$.

\vspace*{-1.9cm}
\hspace*{-1.8cm}
\medskip
\epsfxsize=12.5truecm
\epsfbox{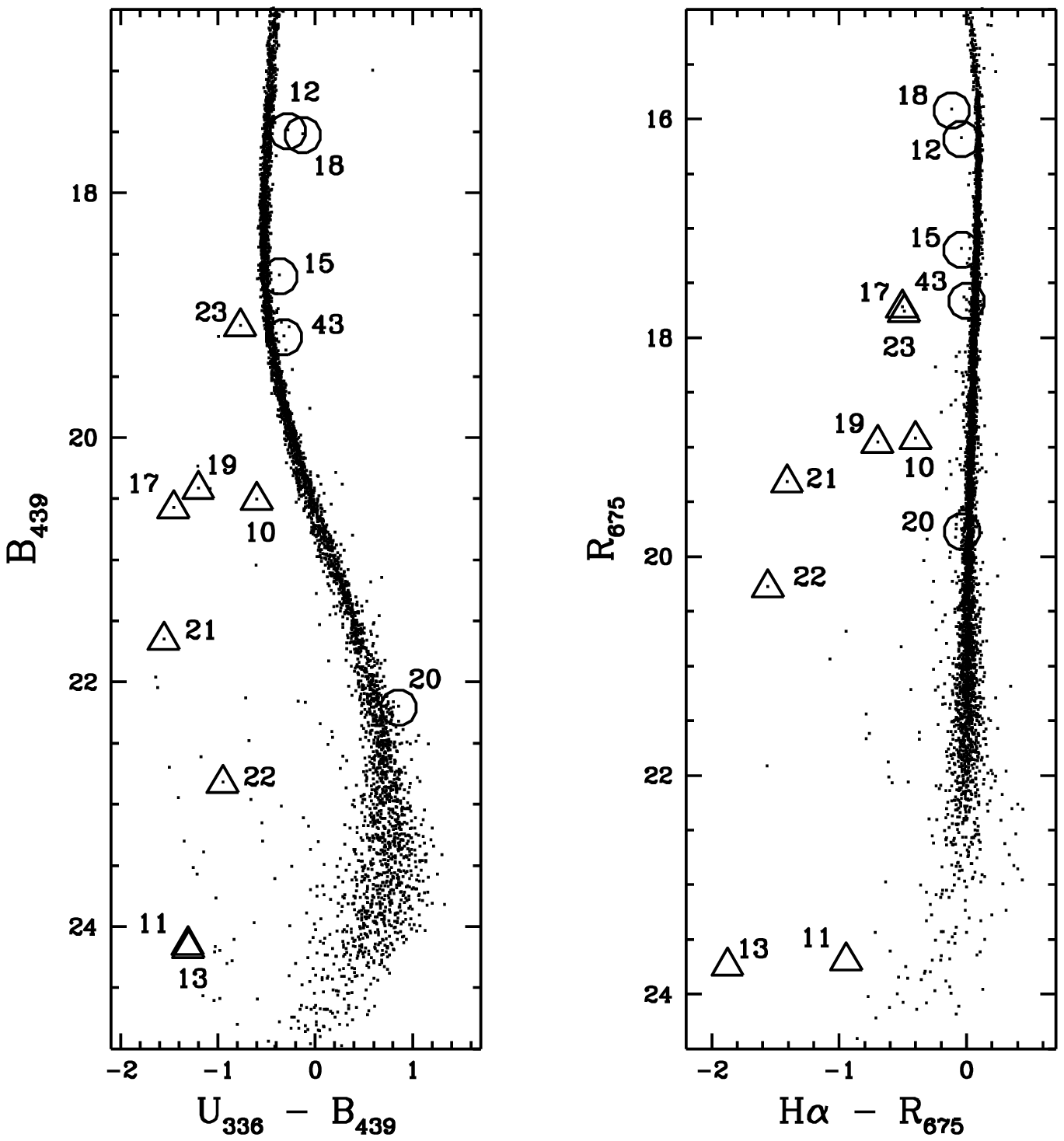}
\vspace*{-1.2cm}
\figcaption[f3.eps]
{Left, HST color magnitude diagrams (cf. Grindlay 
et al, in preparation) for a) (Left) U-B vs. B 
and  b) (Right) \Halpha - R vs. R. Chandra sources identified with 
CVs ($\Delta$) and BY Dra binaries ($\bigodot$) are numbered with 
the source numbers given in Table 1.  
Stars for which membership has been confirmed by proper motion 
(Cool and Bolton 2001) are shown as dots.}
\bigskip

Source U12, optically identified by TGEC with a BY star (WF4-1), is 
in fact the eclipsing binary MSP 
(PSRJ1740-5340) previously discovered in NGC 6397  
(D'Amico et al 2001a) for which the 
pulsar timing position (D'Amico et al 2001b) suggests the same optical 
counterpart (Ferraro et al 2001). The offset in position, 
(radio-x-ray) $\delta$(RA,Dec)= +0.16\arcsec, -0.94\arcsec, is 
consistent with the Chandra astrometric uncertainties. 
The BY signatures (main sequence like 
binary; weak \Halpha emission) are likely due to 
MSP heating (or shock excitation) of the main sequence companion 
in this unusual long-period (1.35d) binary system, which 
is evident in the x-ray  variation of U12: the smooth rise 
in x-ray flux is consistent with the phase of egress from 
radio eclipse. Further details, and comparison with the MSPs 
detected by Chandra 
in 47Tuc (GHEM), are given in Grindlay et al, in preparation.

Surprisingly, none of the ``non-flickerers'' (Cool et al 1998, TGEC) 
which we have identified as containing 
He White Dwarfs (Edmonds et al 1999), are detected with Chandra. 
The suspected ROSAT detection of NF3 (Verbunt and Johnston 2000)  
is in fact U18, identified here with a neighboring BY Dra 
candidate 1.1\arcsec\ away, and similar to the MSP U12 in its position 
in the CMD (Fig. 3). 
This suggests the HeWDs in NGC 6397, with radial distribution 
indicating they contain ``dark'' binary companions (TGEC), 
do not contain neutron stars which would likely be 
detectable as MSPs. Rather, as suspected on evolutionary 
grounds, the primaries are probably CO WDs (Hansen et al 2001).

\section{DISCUSSION AND CONCLUSIONS}
Comparison of Figure 1a and Table 1 reveals that the unidentified 
sources are all (except U25 and U61, both of which have 
possible faint blue counterparts (GTEC)) at \gsim1\arcmin from 
the cluster center. Chandra deep surveys (Giacconi et al 2001) 
predict \about5 AGN background sources in this 
exposure out to 2\arcmin.  
Indeed, our study of the radial source distributions 
measured with Chandra and HST (GTEC) shows that 
the Chandra sources are fit by a King model with 
core radius r$_c$ \about6\arcsec and a power law index slope 
of -2.5 at larger radii, with 20 cluster sources 
within 2\arcmin (\about20 core radii) and 
\Lx \gsim5 \X 10$^{29}$\lcgs (or 15 sources at 
\Lx \gsim10$^{30}$\lcgs). In contrast, 47Tuc 
contains 187 sources within 4\arcmin (\about16 x-ray 
core radii) and \Lx \gsim10$^{30}$\lcgs (Grindlay et 
al, in preparation). Although the \about1:12 ratio of 
x-ray source numbers in NGC 6397 vs. 47Tuc scales 
nearly with cluster mass (\about1:6, cf. Pryor and 
Meylan 1993), the 
source population of NGC 6397 is relatively 
devoid of the faint, soft sources in the lower right of the 
x-ray CMD (Fig. 2) that contain most of the MSPs in 47Tuc. 
Indeed, the one MSP in NGC 6397 now detected in x-rays (U12) 
has Xcolor similar to the ``bluest'' MSP (J) in 47Tuc. 
Possible ``red'' MSP candidates in the central regions of 
NGC 6397 are U61 and U41, as well as a soft source just 
visible in the wings of the bright CV U17 
(cf. Fig. 1b) at RA, Dec 
42.72(3), 40:21.1(5) in the notation of Table 1. Crowding 
prevented its detection by Wavdetect, but we estimate 
medcts \about10 (\Lx \about1 \X 10$^{30}$) 
and Xcolor \gsim1.3 from examination of the image. 
U5, U7, U31 and U42 are also all possible MSPs, 
though at such large offsets from the core they 
would have to have been ejected (the rough co-alignment of sources 
between U7 and U28 with the line of 5 bright blue stragglers in 
the cluster core (Auriere, Ortolani and Lauzeral 1990) 
may suggest preferential cluster rotation 
and binary ejection; U12 is along this ``line'', as expected 
if it was ejected in an exchange encounter). We estimate 
a total of perhaps 5 MSPs in NGC 6397 if the Xcolor distribution 
from 47Tuc (GHEM) is a guide. 

The numbers of detected (radio) vs. suspected (x-ray) 
MSPs in NGC 6397 (1 vs. \about5) compared to 47Tuc (20 vs. \about100)  
are similar in ratio (1/5; which may indicate radio vs. x-ray beaming 
factors) but scale a factor of \about3 below that for cluster mass. 
In  contrast, NGC 6397 appears to have \about10 CVs 
whereas 47Tuc has \about20-30. Thus the two clusters suggest a 
dichotomy of compact objects and binary production: 
NGC 6397 has over-produced CVs (for its mass), perhaps due 
to 2-body captures during core collapse. Its relatively lower 
MSP production might be due to a lower NS 
retention or fraction due to differences in the cluster initial mass 
functions in turn due to their very different metallicities. 
The  populations of BY Dra systems detected 
in x-rays in NGC 6397 vs. 47Tuc 
(4 vs. 6 Chandra sources, respectively, but probably 
\gsim100 in 47Tuc when crowding and the factor of \about2 
higher \Lx limits are considered) is consistent with the 
depletion of main sequence binaries in the core of 
NGC 6397 (Cool and Bolton 2001). However the identification 
of U12 as in fact a MSP raises the possibility that a 
number of BY candidates in 47Tuc are in fact wide binary 
MSP systems, resulting from double exchange collisions, but which 
(unlike U12) have sunk back into  the more massive cluster core. 

\acknowledgments

We thank Haldan Cohn, Phyllis Lugger and Jacob Taylor 
for discussions and Don Lloyd for his H and He-atmosphere models. 
This observation was conducted as part of the Chandra GTO program 
with support from NASA (HRC contract NAS8-38248)

\begin{deluxetable}{lccccccr}
\tablecaption{NGC 6397 Chandra Sources}
\footnotesize
\tablewidth{6truein}
\tablehead{
\colhead{Src} & 
\colhead{RA} &
\colhead{Dec}&
\colhead{Medcts} &
\colhead{kT} &
\colhead{$N_H$} &
\colhead{log(\Lx)} &
\colhead{ID}\\
 & (sec) & (\arcmin:\arcsec) & (0.5-4.5 keV) & (keV )& ($10^{21}$\cmsq) & 
(\lcgs) & }
\startdata
U5    &   54.51(1)     & 40:44.9(1)    &    18  &  --     &  -  & 30.2 &  \\
U7    &   52.807(8)    & 41:21.89(6)   &    54  &  --     &  -  & 30.7 &  \\
U10   &   48.948(4)    & 39:48.86(3)   &   215  &  7-200  & 12-22 & 31.7 & CV 
6 \\
U11   &   45.743(9)    & 40:41.99(9)   &    17  &  --     &  -  & 30.2 & CV 7 
\\
U12   &   44.571(6)   & 40:41.84(4)   & 66  & 1-7 & 0.9-6  & 30.9 &
MSP \\
U13   &   44.054(7)    & 40:39.55(6)   &    34  &  --     &  -  & 30.5 & CV 8 
\\
U14   &   43.26(1)     & 41:55.46(6)   &    12  &  --     &  -  & 30.0 &  \\
U15   &   42.873(7)    & 40:34.12(9)  &  28  &  --   &  -  & 30.4 & BY(PC-2) \\
U16   &   42.61(1)     & 42:15.8(1)    &     8  &  --     &  -  & 29.9 &  \\
U17   &   42.607(1)    & 40:19.62(1)   &  2243  &  6-10   & 1.6-2.3 & 32.3 & 
CV 3 \\
U18   &   42.567(3)    & 40:27.93(2)   &   277  &  33-200 & 0-1.6  & 31.3 & BY 
\\
U19   &   42.258(2)    & 40:29.03(1)   &   881  &  11-91  & 2.5-4 & 31.9 & CV 
2 \\
U20   &   41.98(1)     & 40:37.8(1)    &     6  &  --     &  -  & 29.7 & BY \\
U21   &   41.794(4)    & 40:21.66(4)   &   188  &  5-120  & 2.1-4 & 31.3 & CV 
4 \\
U22   &   41.659(2)    & 40:29.37(1)   &  1019  &  9-30   & 1.1-2.0 & 31.9 & 
CV 5 \\
U23   &   41.551(1)    & 40:19.59(1)   &  1217  &  6-11   & 4.5-6 & 32.2 & CV 
1 \\
U24   &   41.421(2)    & 40: 4.73(2)   &   640  & 0.057-.092 & 1-2.6 & 31.9 & 
qLMXB\\
U25   &   41.20(1)     & 40:26.3(1)    &    13  &  --     &  -  & 30.1 &  \\
U28   &   38.860(3)    & 39:51.45(3)   &   189  &  6-200  & 2.7-7 & 31.3 & CV 
\\
U31   &   34.21(1)     & 41:15.8(1)    &     7  &  --     &  -  & 29.8 &  \\
U41   &   44.97(1)     & 39:55.6(1)    &     6  &  --     &  -  & 29.7 &  \\
U42   &   43.01(1)     & 38:31.3(1)    &    18  &  --     &  -  & 30.2 &  \\
U43   &   40.48(1)     & 40:23.0(1)    &     3  &  --  &  -  & 29.4& BY(PC-4)\\
U60   &   47.82(3)     & 41:27.7(1)    &     3  &  --     &  -  & 29.4 &  \\
U61   &   45.19(2)     & 40:29.1(2)    &     7  &  --     &  -  & 29.8 &  \\
\enddata

\tablecomments{Table columns: 1. Wavdetect source number.
2,3. RA in time seconds, added to 17:40; Dec in 
\arcmin and \arcsec, added to -53\deg; epoch J2000. 
4. Counts detected by Wavdetect (in CIAO software,
available at http://asc.harvard.edu) in medium (0.5-4.5 keV) band.
5. Best fit bremsstrahlung temp (CVs) 90\% confidence
interval, allowing kT, $N_H$, normalization free.  For U24, H atm
kT$_{eff}$,  90\% confidence.
6. Best $N_H$ column fit with bremsstrahlung (or H atm) spectrum,
in units of $1\times10^{21}$ cm$^{-2}$, the fiducial cluster value.
7. Unabsorbed \Lx values (in \lcgs) for 0.5-2.5 keV band, using
the listed (column 5) best fit bremsstrahlung or NS spectrum, or assuming
1 keV bremss spectrum, $N_H$=1$\times10^{21}$\cmsq, 2.5 kpc distance, and thus 
\Lx
=9.3$\times10^{28}$\lcgs ct$^{-1}$. 
8. Tentative source ID:  BY for BY Draconis
main-sequence binaries (IDs from TGEC), CV for cataclysmic 
variables (see text for numbering), 
MSP for millisecond pulsar and qLMXB for quiescent low-mass x-ray
binary.}

\end{deluxetable}


\begin{references}

\reference{Aldcroft2000}
{Aldcroft}, T.~L., {Karovska}, M.,
{Cresitello-Ditmar}, M.~L., {Cameron}, R.~A., {Markevitch},
M.~L. 2000, Proc. SPIE, 4012, 650
 
\reference{Auriere90}Auriere, M., Ortolani, S. and Lauzeral, C. 
1990, Nature, 344, 638

\reference{Brown98}
{Brown}, E.~F.,{Bildsten}, L., \& {Rutledge}, R.~E. 1998,  ApJ 504,
L95 

\reference{Cool93}Cool, A.M., Grindlay, J.~E., Krockenberger, M., 
and Bailyn, C.~D. 1993, ApJ, 410, L103

\reference{Cool01}Cool, A.~M., and Bolton, A.~S. 2001, in 
"Stellar Collisions and Mergers," ASP Conf. Series, ed. M. Shara

\reference{Cool95}Cool, A., Grindlay, J., Cohn, H., Lugger, P., 
and Slavin, S. 1995, ApJ, 439, 695

\reference{Cool98}Cool, A.~M., Grindlay, J.~E., Cohn, H.~N., Lugger, P.~M.,
and Bailyn, C.~D. 1998, ApJ, 508, L75

\reference{Damico01a}D'Amico, N., Lyne, A., Manchester, R., Possenti,
A. and Camilo, F. 2001a, ApJ, 548, L171

\reference{Damico01b}D'Amico, N., Possenti, A., Manchester, R., 
Sarkissian, J., Lyne, A. and Camilo, F. 2001b, ApJ, submitted

\reference{Edmonds}Edmonds, P., Grindlay, J., Cool, A.,  
Cohn, H., Lugger, and Bailyn, C. 1999, ApJ, 516, 250

\reference{Ferraro}Ferraro, F., Possenti, A., D'Amico, N. and 
Sabbi, E. 2001, ApJ, submitted

\reference{Giacconi}Giacconi, R. et al 2001, ApJ, 551, 624

\reference{Grindlay99}Grindlay, J.~E. 1999, in "Annapolis Workshop on Magnetic
Cataclysmic Variables," ASP Conf. Series, Vol. 157, eds. 
C. Hellier and K. Mukai, p. 377

\reference{Grindlay95}Grindlay, J., Cool, A., Callanan, P., 
Bailyn, C., Cohn, H., and Lugger, P. 1995, ApJ, 455, L47  

\reference{Grindlay01}
{Grindlay}, J.~E., {Heinke}, C.~O., {Edmonds}, P.~D., \& {Murray},
S.~S. 2001,  Science 292, 2290 (GHEM)


\reference{Hansen01}Hansen, B., Kalogera, V., Phahl, E. and Rasio, F. 
2001, ApJ, in press

\reference{Heinke01}
Heinke, C.~O., Grindlay, J.~E., Lloyd, D.~A., \& Edmonds, P.~D. 2001, ApJ 
submitted (HGLE)

\reference{Lattimer01}
{Lattimer}, J.~M. \& {Prakash}, M. 2001,  ApJ 550, 426

\reference{Pryor}Pryor, C. and Meylan, G. 1993, in ``Structure and 
Dynamics of Globular Clusters'', ASP Conf. Series, Vol. 50, 
eds. S. Djorgovski and G. Meylan, p. 357

\reference{Rajagopal96}
{Rajagopal}, M. \& {Romani}, R.~W. 1996,  ApJ 461, 327

\reference{Rutledge01}
{Rutledge}, R.~E., {Bildsten}, L., {Brown}, E.~F., {Pavlov},
G.~G., \& {Zavlin}, V.~E. 2001,  ApJ, submitted 

\reference{Sosin97}Sosin, C. 1997, Ph.D. Thesis, University of California

\reference{Taylor01}Taylor, J. M., Grindlay, J. E., Edmonds, P. D., 
Cool, A. M. 2001, ApJ, 553, 169 (TGEC)

\reference{Verbunt00}Verbunt, F. and Johnston, H. 2000, A\&A, 358, 910

\reference{Zavlin96}
{Zavlin}, V.~E., {Pavlov},
G.~G. \& {Shibanov}, Yu.~A. 1996,  A\&A, 315, 141

\end{references}
\end{document}